\documentclass{phbauth}
\usepackage{graphicx}
\begin{document}
\begin{frontmatter}
\title{Quantum Hole Digging in Magnetic Molecular Clusters}

\author[address1]{W. Wernsdorfer\thanksref{thank1}},
\author[address2]{T. Ohm}
\author[address3]{C. Sangregorio}
\author[address3]{R. Sessoli}
\author[address3]{D. Gatteschi}
\author[address2]{C. Paulsen}

\address[address1]{Lab. L. N\'eel - CNRS, BP166, 38042 Grenoble, France}
\address[address2]{CRTBT - CNRS, BP166, 38042 Grenoble, France}
\address[address3]{Dept. of Chem., Univ. of Florence, 50144 Firenze, Italy}

\thanks[thank1]{Corresponding author. E-mail: wernsdor@labs.polycnrs-gre.fr} 

\begin{abstract}
Below 360~mK, Fe$_8$ magnetic molecular clusters are in the pure 
quantum relaxation regime. We showed recently that the predicted 
``square-root time'' relaxation is obeyed, allowing us to develop a 
new method for watching the evolution of the distribution of molecular 
spin states in the sample. We measured the distribution $P(H)$ of molecules 
which are in resonance at the applied field $H$. Tunnelling initially causes rapid 
transitions of molecules, thereby ``digging a hole'' in $P(H)$. For 
small initial magnetisation values, the hole width shows an intrinsic 
broadening which may be due to nuclear spins. We present here hole digging measurements in the thermal activated regime which may allow to study 
the effect of spin-phonon coupling.
\end{abstract}

\end{frontmatter}

Magnetic molecular clusters such as Mn$_{12}$ac and Fe$_8$ are
ideal systems to study quantum tunnelling 
of the magnetisation (QTM) \cite{Sessoli98}. 
Crystals of these materials can be thought of as ensembles of 
identical, iso-oriented nanomagnets of net spin $S$~= 10, and with a strong Ising-like anisotropy. 
Theoretical discussion of 
thermally-activated QTM assumes that thermal processes (principally 
phonons) promote the molecules up to high levels, not far below the 
top of the energy barrier, and the molecules then tunnel inelastically 
to the other side.

At temperatures below 360~mK, Fe$_8$ molecular clusters display a 
clear crossover from thermally activated relaxation to a temperature 
independent quantum regime \cite{Sangregorio97}. In this regime only the two lowest 
levels of each molecule are occupied, and only ``pure'' quantum 
tunnelling through the anisotropy barrier can cause direct transitions 
between these two states. 

Recently, we developed a method \cite{Ohm98,Wernsdorfer99} for measuring 
the intrinsic line width broadening due to local fluctuating fields of the nuclear spins. It is 
based on the general idea that the short time relaxation rate is directly connected to the 
number of molecules which are in resonance at a given longitudinal applied field $H$. 
In the low temperature regime, the Prokof'ev - Stamp theory \cite{Prokofev98} predicts that the magnetisation should relax at short times with a square-root time dependence, giving the rate function $\Gamma_{\rm sqrt}(H)$ which is
proportional to the normalised distribution $P(H)$ of molecules which are in resonance at 
the applied field $H$.

\begin{figure}[t]
\begin{center}\leavevmode
\includegraphics[width=0.8\linewidth]{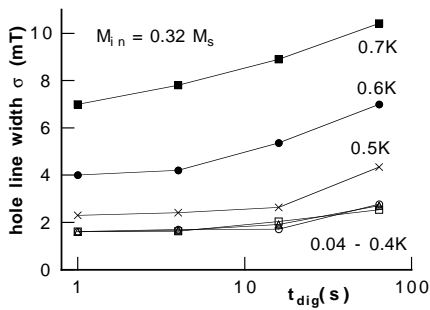}
\caption{Digging time and temperature dependence of the hole line width $\sigma$.}
\label{fig1}
\end{center}
\end{figure}

Our measuring procedure is as follows. Starting from a well defined magnetisation state, we apply a magnetic field $H$ in order to measure the short-time relaxation behaviour, yielding the rate function $\Gamma_{\rm sqrt}(H)$ at the field $H$. Then, starting again from the same well defined magnetisation 
state, we measure $\Gamma_{\rm sqrt}(H)$ at another field $H$, yielding the field 
dependence of $\Gamma_{\rm sqrt}(H)$ which is proportional to the dipolar distribution 
$P(H)$.

This technique can be used for following the time evolution of molecular states in 
the sample during a tunnelling relaxation \cite{Wernsdorfer99}. Starting from a well 
defined magnetisation state, and after applying a field $H_{dig}$, we let 
the sample relax for a time $t_{dig}$, called 'digging field and digging time', 
respectively. During the digging time, a small fraction of the molecular spins tunnel and 
reverse the direction of their magnetisation. Finally, we apply a field $H$ to measure the 
short time relaxation in order to get $\Gamma_{\rm sqrt}(H)$. The entire 
procedure is then repeated to probe the distribution at other fields $H$ yielding 
$\Gamma_{\rm sqrt}(H,H_{dig},t_{dig})$ which is proportional to the number of spins 
which are still free for tunnelling.

We used this hole digging' method, 
for studying Fe$_8$ and Mn$_{12}$ac molecular clusters \cite{Wernsdorfer99} and found that tunnelling 
causes rapid transitions of molecules near $H_{dig}$, thereby "digging a hole" in 
$P(H,H_{dig},t_{dig})$ around $H_{dig}$, and also pushing other molecules away 
from resonance. The hole widens and moves with time, in a way depending on sample 
shape; the width dramatically depends on thermal annealing of the magnetisation of the 
sample. For small initial magnetisation, the hole width shows an intrinsic 
broadening which may be due to nuclear spins \cite{Wernsdorfer99}. The hole could be fitted to a Lorentzian 
function yielding the line width $\sigma$ which we studied as a function of 
temperature and digging time (fig. 1). We defined an intrinsic line width $\sigma_0$ by a linear 
extrapolation of the curves to $t_{dig}$ = 0 (fig. 2). For temperatures between 0.04 and 0.4K, $\sigma_0$ $\approx$ 1.6 mT. For T $>$ 0.4 K, $\sigma_0$ increase rapidly.

The physical origin of the line width $\sigma_0$ at T $<$ 0.4 K is assigned to the fluctuating hyperfine fields \cite{Prokofev98}. At higher temperature, spin-phonon coupling might be responsible for the observed line width.

\begin{figure}[t]
\begin{center}\leavevmode
\includegraphics[width=0.8\linewidth]{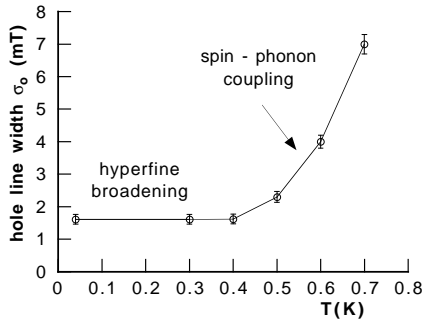}
\caption{Temperature dependence of the intrinsic hole linewidth $\sigma_0$ of molecular clusters Fe$_8$.}
\label{fig2}
\end{center}
\end{figure}

%

\end{document}